\documentstyle[11pt]{article}
\newcommand{\ba}{\begin{eqnarray}}
\newcommand{\ea}{\end{eqnarray}}
\newcommand{\be}{\begin{equation}}
\newcommand{\ee}{\end{equation}}
\newcommand{\nn}{\nonumber \\}
\newcommand{\vk}{{\bf{k}}}

\newcommand{\vx}{{\bf{x}}}
\newcommand{\ls}{\mathrel{\raise1.16pt\hbox{$<$}\kern-7.0pt 
\lower3.06pt\hbox{{$\scriptstyle \sim$}}}}         
\newcommand{\gs}{\mathrel{\raise1.16pt\hbox{$>$}\kern-7.0pt 
\lower3.06pt\hbox{{$\scriptstyle \sim$}}}}         

\long\def\comment#1{}

\def\fun#1#2{\lower3.6pt\vbox{\baselineskip0pt\lineskip.9pt
  \ialign{$\mathsurround=0pt#1\hfil##\hfil$\crcr#2\crcr\sim\crcr}}}
\def\lap{\mathrel{\mathpalette\fun <}}
\def\gap{\mathrel{\mathpalette\fun >}}

\begin{document}
\title{Non-gaussianity vs. non-linearity of cosmological perturbations}
\author{Licia Verde\\
\small{IfA,University of Edinburgh}\\
\small{Royal Observatory, Blackford Hill, EH9 3HJ, Edinburgh, U.K.}\\
\small{Ph:+44-131-6688393; Fax:+44-131-6688416; lv@roe.ac.uk}}
\maketitle
\begin{abstract}
Following the discovery of the cosmic microwave background, the hot
big-bang model has become the standard cosmological model.  In this theory,
small primordial fluctuations are subsequently amplified by gravity to form
the large-scale structure seen today. Different theories for unified models of
particle physics, lead to different predictions for the statistical properties
of the primordial fluctuations, that can be divided in two classes: gaussian
and non-gaussian.  Convincing evidence against or for gaussian initial
conditions would rule out many scenarios and point us towards a physical
theory for the origin of structures.

The statistical distribution of cosmological perturbations, as we
observe them, can deviate from the gaussian distribution in several different
ways. Even if perturbations start off gaussian, non-linear gravitational
evolution can introduce non-gaussian features. 
Additionally, our knowledge of the Universe comes principally from the study of
luminous material such as galaxies, but galaxies might not be faithful tracers
of the underlying mass distribution. The relationship between fluctuations in
the mass and  in the galaxies distribution ({\it bias}), is often assumed to be
local, but could well be non-linear.
Moreover, galaxy catalogues use the redshift as third spatial coordinate:
the resulting redshift-space map of the galaxy distribution is non-linearly
distorted by peculiar velocities.
Non-linear gravitational evolution, biasing, and redshift-space distortion
introduce non-gaussianity, even in an initially gaussian fluctuation field.

I will investigate the statistical tools that allow us, in principle, to
disentangle the  above different effects, and the observational datasets we
require to do so in practice.
\end{abstract}

\section{Introduction}
Until recently in cosmology, non-gaussianity has been a synonymous of
non-linearity; but, in the last 5 years or so, more and more objects like the
galaxy of [1] at redshift 5.6 have been found.
For the first time a galaxy has been found at higher redshift than the most
distant known quasar. More recently, a galaxy at redshift almost 7 has been found [2]. 
The standard ``inflationary'' cosmological model with gaussian initial
conditions predicts that these objects should be very rare.  
It is becoming increasingly difficult to accommodate the existence of so many
high-redshift galaxies under the assumption that non-gaussianity is equivalent
to non-linearity, that is postulating gaussian initial conditions.
Non-gaussianity does not necessarily imply non-linearity: there might be
some primordial non-gaussianity and it is necessary to ``find a way'' to
distinguish the two effects.

\section{Non-gaussianity due to non-linearities}
Let us start by assuming gaussian initial conditions and investigate the
effects of non-linearities.
We define the fractional density contrast $\delta$ as $\delta \rho/\rho$,
where $\rho$ is the mean density. The
probability distribution of $\delta$ starts off symmetric around zero, with
negligible tails for $|\delta|>1$. 
Non-linear gravitational evolution skews the distribution  towards high
densities: this is due to the fact that underdense regions cannot become more
empty than the void ($\delta\geq-1$) while overdense regions can accrete
matter arbitrarily (no upper limit on $\delta$).
This is not the only process that can skew an initially gaussian
distribution. The mass in the Universe is mainly dark
matter and cannot be observed directly: only galaxies can be observed, but
mass and galaxy distributions may not be identical: the idea that galaxies are
biased tracers of the mass distribution  was introduced in the early eighties
[3]\footnote{Although the fact that galaxies of different morphologies have different spatial
distributions and they
cannot all be  good tracers of the underlying mass distribution, was known much before the introduction of the concept of  bias
(e.g. [4]).} and has featured strongly in large scale structure (LSS)
studies. 
In general, bias must alter the statistics of any underlying matter
distribution, otherwise $\delta < - 1$ for the galaxy field, which corresponds to a negative galaxy
density. In different bias schemes suggested in the literature, the relation between the galaxy and the mass
fluctuation fields ($\delta_g$ and $\delta$ respectively) has been taken to be
local, non-local, eulerian, lagrangian, stochastic etc... 

In what follows we will assume that $\delta_g(\vx)=F[\delta(\vx)]$, that is the
bias, is a local eulerian function of the underlying mass field. Furthermore we will
assume (following [5]) that $F$ can be expanded in Taylor
series and we will truncate the expansion to the quadratic term:
\be
\delta_g(\vx)=b_0+b_1\delta(\vx)+\frac{b_2}{2}\delta^2(\vx)+\ldots
\label{eq:nlbias}
\ee
$b_0$ is unimportant and simply ensures that $\langle\delta_g\rangle=0$.
This non-linear operation on the matter field introduces some skewness,
i.e. some non-gaussianity. 
As first suggested by Fry [6] it is possible to disentangle the two
non-gaussian contributions, non-linear gravity and bias, by looking at higher-order correlations in the mildly non-linear regime. In particular,
if the initial fluctuations are gaussian and cosmological structures grow by
gravitational instability, the three-point correlation function is
intrinsically a second-order quantity\footnote{In the quantity $\delta$,
assumed to be small.} and is detectable in the mildly
non-linear regime. If then bias can be expressed as in
equation (\ref{eq:nlbias}), 
it is possible to show that a likelihood analysis\footnote{The likelihood method can easily be generalized
to measure the lagrangian [7] and stochastic (e.g. [8]) bias parameters.} of the bispectrum (the three-point correlation
function in Fourier space) can yield $b_1$ and $b_2$. 

The bispectrum $B(\vk_1\vk_2\vk_3)$ is defined as 
\be
\langle\delta_{k_1}\delta_{k_2}\delta_{k_3} \rangle=(2\pi)^3B(\vk_1\vk_2\vk_3)\delta^D(\vk_1+\vk_2+\vk_3) 
\ee
where $\delta_k$ is the Fourier transform of $\delta (\vx)$. Due to the presence of the Dirac delta
function $\delta^D$, the bispectrum can be non zero only when the three $\vk$ form a triangle. 

In practice, the higher-order statistic (the bispectrum) exploits the fact
that gravitational instability skews the density field as it evolves, creating
sheets and filament-like structures reminiscent of the Zeldovich pancakes. 
Non-linear
bias also  introduces skewness  but  does so by shifting the iso-density
contours up and down, without modifying the shape of the structures.  These two effects can be
disentangled by using different triangle shapes for the bispectrum.

There are several advantages in performing this sort of analysis in Fourier
space, most of them are the same advantages of the power spectrum over the
two-point correlation function. We will recall here only the following: the estimates of power on different scales can
be made uncorrelated; it is easy to deal with the error estimate; and more
importantly it is easy to
distinguish between linear, mildly non-linear and highly non-linear scales.
In the mildly non-linear regime the bispectrum is given by:
\be
\langle\delta_{g,\vk_1}\delta_{g,\vk_2}\delta_{g,\vk_3} \rangle\!=\!(2\pi)^3\!
P_g(k_1)P_g(k_2)\!\left[c_1J(\vk_1,\vk_2)\!+\!c_2 \right]\delta^D\!(\vk_1\!+\vk_2\!+\vk_3\!)\!+\!cyc.
\label{eq:bisp1}
\ee
where $P_g$ denotes the galaxy power spectrum, $J(\vk_1,\vk_2)$ is a known
function of the two $k$-vectors and
\be
c_1=\frac{1}{b_1} \;\;\;\;\;\;c_2=\frac{b_2}{b_1^2}\;.
\label{eq:c1c2}
\ee
In the absence of bias ($c_1=1$, $c_2=0$) it would be easy to isolate the non-gaussianity
generated by gravitational instability. Nevertheless, even in the presence of
bias, it is possible to disentangle gravity from bias.
Equation (\ref{eq:bisp1}) is in a form suitable for a likelihood analysis for
the two bias parameters via $c_1$ and $c_2$ [9], once the covariance is
known\footnote{The bispectrum is a three point quantity, its covariance is a
six-point quantity (pentaspectrum).}. 
A generating functional approach to calculate {\it analytically} the N-point function and
therefore the covariance for the bispectrum was introduced in [10,11].
The performance of the method has been tested on  biased and unbiased N-body
simulations\footnote{The N-body simulation was provided by the
Hydra-consortium and produced using the code [12].}, with very promising results\footnote{For a different approach
see J. Frieman contribution in this volume.} [11] for the application to
forthcoming galaxy redshift surveys such as SDSS and 2dF. An estimation for
the expected error achievable from present galaxy surveys, yields an error on
$c_1$ of about 100\% [11] and is therefore not particularly useful. 

\section{Real world issues} 
Of course, reality is always more complicated: in a realistic galaxy survey
several complications arise due to the presence of shot noise and selection
function, but more importantly due to redshift space distortions.
Galaxy surveys in fact use the redshift as the third spatial coordinate. The
redshift would be an accurate distance indicator in a perfectly homogeneous
Universe; but the universe is clumpy, inhomogeneities perturb the Hubble flow
and introduce peculiar velocities. The resulting redshift-space map of the
galaxy distribution is thus distorted along the line of sight, and the nature
of this distortion is intrinsically non-linear.
On large scales the coherent inflow into overdense regions introduces  a
squashing effect in the  redshift map (the {\it great wall} effect), on
smaller scales, the virialized highly non-linear structures appear elongated
along the line of sight (the {\it fingers-of-God}), heavily contaminating the
mildly non-linear regime were most of the
signal for the bispectrum comes from. 
In [13] we showed that, with an accurate modeling of redshift space
distortions in the distant observer approximation, it is possible to disentangle the effects
of redshift-space anisotropies from the bias and gravitational
effects. This is achieved by combining a second-order perturbation theory
description of the coherent inflow (e.g. [14])
with an exponential velocity dispersion model, and discarding the $k$-modes 
where the contamination from highly non-linear structures is too big to be
successfully modeled.
An alternative approach hss also been explored (e.g. [15]).

\begin{figure}[h]
\begin{center}
\begin{picture}(160,160)
\includegraphics{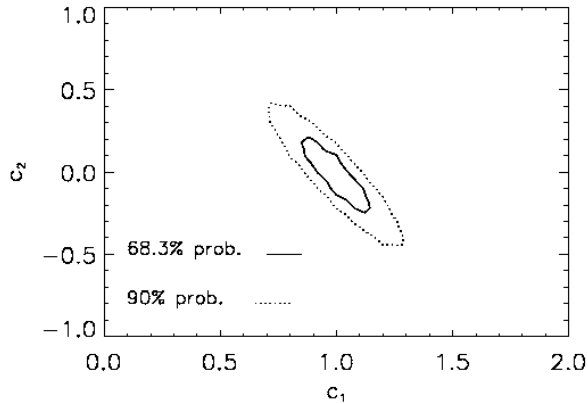}
\end{picture}
\label{fig:redbias}
\caption[]{{\it Likelihood for $c_1=1/b_2$ and $c_2=b_2/b_1^2$,
(where $b_1$ and $b_2$ are linear and quadratic bias parameters) for an
unbiased simulation ($b_1=1$, $b_2=0$) in redshift space. The solid and dotted
lines show the 1 and 3 $\sigma$ confidence levels respectively. This figure shows
that it is possible to disentangle the non-linear
gravitational instability, bias and redshift space distortion signals, and
also that  from future galaxy redshift surveys such as SDSS and 2dF, the bias
could be known with an accuracy better that 10\%.}}
\end{center}
\end{figure}

The result of the likelihood analysis performed on an redshift space unbiased
simulation of 100$h^{-1}$ Mpc side is shown in figure 1. 

It shows not only that it is possible to disentangle the non-linear
gravitational instability, bias and redshift space distortion signals, but
also that from future galaxy redshift surveys such as Sloan digital sky survey
(SDSS) and the Anglo-Australian two-degree field (2dF), the bias
could be known with an accuracy better that 10\%. 
Before achieving this goal however, there are several other issues to deal
with, that are potentially serious for any Fourier based technique. In
particular,  these are the mask, the
difficulty of obtaining redshifts for close pairs of galaxies, the
holes arising from bright-star drills, and the variable completeness.   
We have investigated and quantified these effects on a simulated catalogue
and concluded that a 10\% error on the bias parameter could be achieved from
the 2dF survey [16].

\subsection{Bypassing the redshift-space distortions}
Two-dimensional
surveys avoid the redshift-space distortion problems, but, in principle, contain less
information. However, because of the smaller observational effort required,
these can contain a much larger number of objects. For example the APM survey
at present contains $10^6$ galaxies, the DPOSS catalogue will have 50 million
galaxies and the SDSS will provide us with a two-dimensional map of $10^7$
galaxies.

To treat the projection of higher-order correlations on the celestial sphere,
the spherical nature of the distribution cannot be ignored. Spherical harmonics are eigenfunctions for the two-dimensional surface of the
sphere and therefore are the natural basis describing a two-dimensional random
field on the sky. When comparing with Fourier space analysis we have that 
\be
\delta_{\vk}\longrightarrow a_{\ell}^{m}
\ee
and in particular for the bispectrum 
\be
\delta^D(\vk_1\vk_2\vk_3)\longrightarrow\left(^{\ell_1\;\;\;\ell_2\;\;\;\ell_3}_{m_1 \;\;m_2\;\; m_3} \right)
\ee
where on the RHS we have the three-J symbol.
To perform the same sort of analysis as the one illustrated in sections 2 and 3
 an expression that relates the 3D bispectrum to the projected one in
spherical harmonics is needed:
\be
\langle a_{\ell_1}^{m_1}a_{\ell_2}^{m_2}a_{\ell_3}^{m_3}\rangle=
\ee
$$
\left(^{\ell_1\;\;\ell_2\;\;\ell_3}_{m_1 m_2 m_3}
\right)\left[\frac{1}{\overline{n}}\frac{16}{\pi}\sqrt{\frac{(2\ell_1+1)(2\ell_2+1)(2\ell_3+1)}{(4\pi)^3}}\right.\times
$$
$$\int\! dk_1 dk_2 i^{\ell_1+\ell_2} k_1^2
k_2^2\Psi_{\ell_1}(k_1)\Psi_{\ell_2}(k_2)
%
\!\sum_{\ell \ell_6\ell_7}\!i^{\ell_6+\ell_7}(-1)^{\ell} B_{\ell}(k_1,k_2)(2
\ell_6+1)(2\ell_7+1)\overline{\rho}\times
$$
$$
\int dr r^2 \psi(r)j_{\ell_6}(k_1r)j_{\ell_7}(k_2r)\left.
\left(^{\ell_1\;\;\ell_6\;\;\ell}_{0 \;\;\;0\;\;\; 0} \right)
\left(^{\ell_2\;\;\;\ell_7\;\;\ell}_{0\;\;\;\; 0\;\;\;\; 0} \right)
\left(^{\ell_3\;\;\ell_6\;\;\ell_7}_{0\;\;\;\; 0\;\;\;\; 0} \right)
\left\{^{\ell_1\;\ell_2\;\ell_3}_{\ell_7\;\ell_6\;\,\ell}
\right\}+cyc.\right]
$$
This expression is quite complicated: the derivation and the detailed
explanation of it can be found in [17]. For the purpose of this contribution we only have to notice
that it is an {\it exact} expression relating the spherical harmonics projected
bispectrum $\langle a_{\ell_1}^{m_1}a_{\ell_2}^{m_2}a_{\ell_3}^{m_3}\rangle$
to the 3D bispectrum expressed through its Legendre coefficients
$B_{\ell}(\vk_i\vk_j)$. $ \Psi_{\ell_i}(k_j)$ is a known function of the
selection function, $j_{\ell}$ denotes the spherical Bessel function and
$\{\ldots\}$ denotes the Wigner 6-J symbol.

\begin{figure}[h]
\begin{center}
\begin{picture}(160,160)
\includegraphics{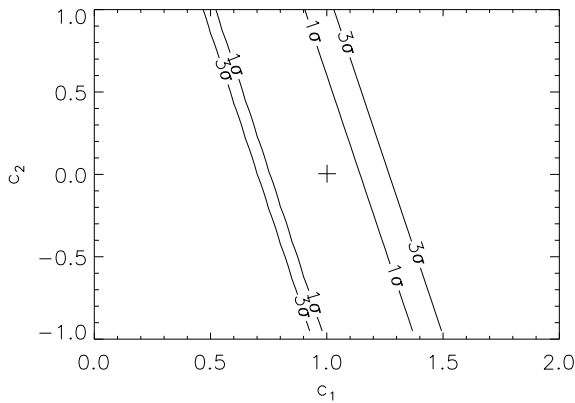}
\end{picture}
\label{fig:projbisp}
\caption[]{{\it Likelihood contours for degenerate triplets
configurations. The two levels are the 1$-\sigma$ and 3-$\sigma$ confidence
levels and the $+$ indicates where the true value for the parameters
lies. Perturbation theory breaks down at $\ell \sim 50$. Adding other
configurations does not help much: this is about the best result obtainable
from projected surveys.}}
\end{center}
\end{figure}

By performing a likelihood analysis to measure the bias parameter on a all sky
simulation with the APM selection function, we find that the results are not
encouraging for projected catalogues (see figure 2).

It is therefore preferable to undertake the bispectrum study of 3D galaxy
redshift surveys such as SDSS and 2dF using the method described above. 
The good news is that the exact expression for the projected bispectrum in
spherical harmonics has applications in a variety of areas such as cosmic
microwave background (CMB) and gravitational lensing studies.

\section{Primordial non-gaussianity: CMB vs. LSS} 
Up to now we have assumed gaussian initial conditions. However, among the
theories for structure formation, only inflation predicts a nearly
gaussian distribution for the primordial fluctuations, with deviations from
gaussianity which are calculable, small and dependent on the specific
inflationary model (e.g. [18,19,20]). In
other  models such as non-standard inflation or topological defects models
initial conditions are non-gaussian.
CMB and LSS data will shortly improve
dramatically: it is therefore timely to ask which of the CMB or LSS will
provide a better probe of the nature of primordial fluctuations.
The advantage of looking at CMB maps is that the fluctuation distribution
should be close to the primeval form, but the disadvantage is that the
amplitude of fluctuations is small and there are foregrounds and other effects
to account for. The advantages of looking at LSS is that the signal has been
amplified by gravity. This is however also a disadvantage because gravity
skews the distribution. Non-linear gravity, bias and redshift space distortions
might completely swamp the primordial signal.  
Following [21], as a discriminating statistic we will use the bispectrum, but we will start by
considering the skewness as an example to illustrate some of the effects.
 
The skewness is defined as 
\be
S_3=\frac{\langle \delta^3\rangle}{\langle\delta^2\rangle^2}.
\ee
For a gaussian field the skewness is zero, while for an initially gaussian field
evolved under gravitational instability to second order in $\delta$, the skewness becomes 34/7 and is constant in
time\footnote{The value 34/7 is strictly true only for an Einstein de Sitter
Universe, but the skewness does not depend strongly on cosmological
parameters.}. In what follows for CMB related calculations, we will assume an
Einstein de Sitter Universe, this assumption is justified because we shall be
concerned with factors of 10 while the cosmology can change the results only by
factors of order unity.
Suppose the initial conditions are very close to gaussian, but with a small
primordial skewness, parameterized by $S_3(z=1100)$ at recombination.
The effect on LSS will be (e.g. [22]):
\be
S_3=S_{3,0}+34/7+{\cal F}
\label{eq:s3}
\ee
where ${\cal F}$ includes a complicated dependence on the three- and four-point
function that we will ignore for the moment and $S_{3,0}$ scales as:
\be
S_{3,0}=\frac{S_3(z)}{(1+z)}\;:
\ee
the primordial skewness redshifts away.
We can then make a thought experiment: assume we know the real space position
of every particle in the whole Hubble volume. The smallest error for the
skewness, that is the smallest $S_{3,0}$ detectable, on 20 $h^{-1}$ Mpc scales
is $S_{3,0}\sim 10^{-2}$, which implies $S_{3}(z=1100)\sim 10$.
We can repeat the exercise for the CMB where now, for consistency, we consider
the smallest detectable skewness on 0.2$^{\circ}$ scales, obtaining
$S_{3}(z=1100)\sim \mbox{few}<10$.

This example already shows that CMB seems to be  more sensitive to primordial
deviation from gaussianity than LSS, but we will
now proceed more  accurately by considering the bispectrum: in fact the bispectrum
contains more information than the skewness and has all the advantages of
being a Fourier space quantity.
In the absence of bias, the LSS bispectrum in second-order perturbation theory for non-gaussian
initial condition is:
\ba
B(\vk_1,\vk_2,\vk_3)& =&
B_0(\vk_1,\vk_2,\vk_3)+2J(\vk_1,\vk_2)P(k_1)P(k_2)+cyc. \nn
 &+& \int d^3kJ(\vk',\vk_3-\vk')T^c(\vk',\vk_3-\vk',\vk_1,\vk_2)+cyc.
\ea 
$B_0$ is the primordial bispectrum evolved linearly and corresponds to
$S_{3,0}$ of equation \ref{eq:s3}; the second term is the usual gravitational
instability bispectrum and correspond to the 34/7 term in equation
\ref{eq:s3}; $T^c$ denotes the Fourier counterpart of the connected four point
function and the integral term correspond to ${\cal F}$ of equation
\ref{eq:s3}.  
We then parameterize the LSS bispectrum as:
\be
B=P(k_1)P(k_2)[2J(\vk_1,\vk_2)c_1+c_2]+cyc.
\ee 
because we know how to estimate $c_1$and $c_2$ from LSS studies (section 2).
In the very idealized case where the real space position of every particle in the
SDSS volume was known, the minimum $c_1$ and $c_2$ detectable would be
respectively  $c_1\sim 10^{-3}$ and $c_2\sim 10^{-2}$ ignoring all the real
world complications of shot noise, selection function etc...

On the CMB side, the bispectrum for realistic non-gaussian models is given by
(e.g. [23]):
\be
B_{\ell_1\ell_2\ell_3}=f(\ell_1,\ell_2,\ell_3)\alpha[C_{\ell_1}C_{\ell_2}+cyc.]
\ee
where $\alpha$ is the amplitude,
$f$ is a known function of the $\ell$ s and $C_{\ell}$ denotes the CMB power spectrum.
The minimum error $\sigma_{\alpha}$ on the amplitude $\alpha$ can be obtained
via the Fisher
information matrix:
\be
\sigma^{-2}_{\alpha}=-\left\langle\frac{\partial^2ln{\cal L}}{\partial{\alpha}^2}\right\rangle\simeq
\sum_{\ell_1\leq\ell_2\leq\ell_3}\frac{(B_{\ell_1\ell_2\ell_3}|\alpha=1)^2}{nC_{\ell_1}C_{\ell_2}C_{\ell_3}}\sum_{m_1m_2m_3}\frac{\left(^{\ell_1\ell_2\ell_3}_{m_1m_2m_3}\right)}{N(m_i,\ell_i)}
\ee
where $n=1/2$ and $N(m_i,\ell_i)$ is the number of non-zero terms like
$C_{\ell_1}C_{\ell_2}C_{\ell_3}$ in the covariance and ranges from 1 to 30.
Here we neglect partial sky coverage effects, and by constraining $\ell\lap
100$ pixel noise and small angular scale effects are negligible.

In [21] we investigated the LSS and CMB bispectrum as a
discriminating statistic for several physically motivated non-gaussian models.
There is an infinitude of deviations from gaussianity and one cannot address
them all: we thus restrict ourselves to physically motivated models
where the non-gaussianity can be dialed from zero (the gaussian limit) and  is
assumed to be small.
In particular we consider the non gaussianity parameterized by:
\be
\Phi=\phi+\epsilon(\phi^2-\langle\phi^2\rangle)
\label{ngmapping}
\ee 
where $\phi$ denotes a gaussian
field and for the moment   
we will assume that $\Phi$ is the gravitational potential; the non-gaussianiy parameter is $\epsilon$, that is zero for a gaussian
field, $\sim 1$ for standard inflation, but can be as big as $\sim 20$ for
some non-standard inflationary models.
The CMB effect is given by $2\epsilon/A_{SW}=\alpha$ where $A_{SW}$ is the
Sachs-Wolfe coefficient $\sim 1/3$. 

It is possible to see [21] that, if the CMB distribution from the future
satellite missions turns out to be  consistent with gaussian, the smallest
$\epsilon$ allowed would be $\sim 20$.

The LSS effect is $c_2=b_2/b_1^2+10^{-6}\epsilon$ and the $T^c$ contribution is negligible.
By substituting the minimum $\epsilon$ measurable from CMB in this expression
 and ignoring bias, we obtain $c_2=10^{-4}$: about two orders of magnitude
smaller that the minimum $c_2$ detectable from LSS even in the most idealized conditions.

In [21] also other physically motivated non-gaussian models have been
considered, the result is qualitatively always the same:
{\it if future CMB maps are consistent with the gaussian hypothesis then any
non-gaussianity seen in the LSS bispectrum is due to non-linear gravity or
bias}, and we know how to disentangle the two.

\section{Looking at smaller scales}
In practice, CMB studies can be affected by noise and foreground and, more importantly, there might be models in which non-gaussianity is present mainly on
LSS or galaxy scales, which are not fully accessible with CMB experiments.
I will therefore investigate another two ways to detect primordial
non-gaussianity on scales smaller than CMB ones.
\subsection{Detecting non-gaussian initial conditions from large-scale structure}
It is possible to  bypass the contamination due to non-linear
clustering and discriminate between gaussian and
non-gaussian initial conditions by using higher-order
statistics in LSS studies such as the trispectrum --that is the connected
four-point correlation function in Fourier space [24]. This quantity has the advantage of having a rather
simple growth rate, with no complicating contributions from non-linear gravity
in second-order
perturbation theory and that the analysis depends on cosmology and bias only
through the measurable quantity\footnote{The quantity $\beta$ arises
naturally when studying the large-scale squashing effect of structures in
redshift-space maps. It is defined as $\beta\simeq\Omega_0^{0.6}/b$ where $b$
is the linear bias parameter.} $\beta$.  The departures from gaussian
statistics can be parameterized by introducing the quantity $H$, which is the
fractional excess of the 4-point function over the gaussian (disconnected)
trispectrum. This quantity, in
specific cases, can give us a meaningful measure of `non-gaussianity': for
mildly non-gaussian fields, a {\em measurement} of $H$ can reliably be made.
For highly non-gaussian fields, the gaussian hypothesis can be rejected, but
the measurement of $H$ will be unreliable.  Following [24] it is possible to deal with
redshift-space distortions, biasing, spatially varying selection function and
shot-noise. 
Figure 3 
shows the minimum $\chi^2$ analysis for the parameter $H$ from a redshift-space
unbiased CDM-like simulation.
\begin{figure}[h]
\begin{center}
\begin{picture}(160,160)
\includegraphics{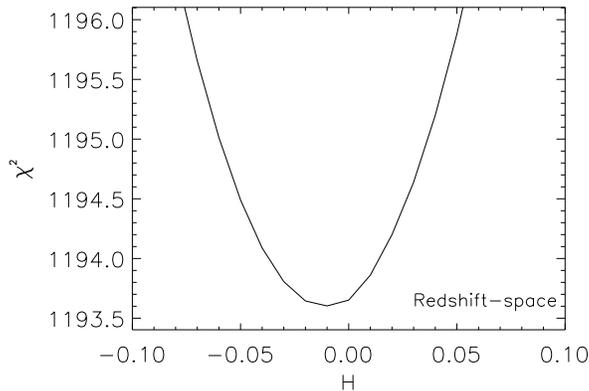}
\end{picture}
\label{fig:trisp}
\caption[]{{\it Minimum $\chi^2$ analysis for the parameter $H$ from a redshift-space
unbiased CDM-like simulation. The analysis is largely bias independent. The
true value for $H$ is 0, and it is nicely within the 1-$\sigma$ level (minimum
$\chi^2$ +0.5.). The quantity $H$ , in
specific cases, can give us a meaningful measure of `non-gaussianity': for
mildly non-gaussian fields.}}
\end{center}
\end{figure}
By applying this method to future galaxy surveys, such as the
SDSS, it will be possible to place tight constraints on
initial departures from gaussian behavior.

\subsection{The abundance of high-redshift objects as a probe of non-gaussian
initial conditions}
LSS probes scales much larger than galaxies but smaller than those accessible
by CMB observations, and  probes the present-day Universe at $z=0$; conversely CMB maps probe the Universe at redshift
$z\sim 1100$ and even larger scales.
The abundance of cosmological structures at redshift in between these two ends
and in particular at $z>1$,
contains also vital information about the nature of
primordial fluctuations due to the fact that one is probing the tail of the distribution.
Figure 4 
shows that the effect of a small non-gaussianity is dramatic on the
 tails: high peaks are amplified much more than low ones and deep through can
become local maxima.
\begin{figure}[h]
\begin{center}
\begin{picture}(160,160)
\includegraphics{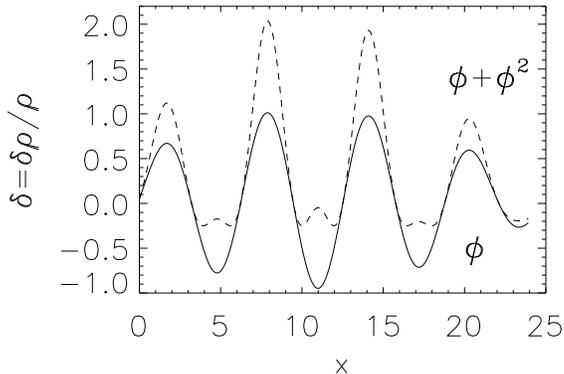}
\end{picture}
\label{fig:sergeiexample}
\caption[]{{\it The effect of a small non-gaussianity are dramatic on the tail of the
distribution. Assume $\phi$ here denotes a one dimensional gaussian
fluctuation filed (black line); the dashed line is given by $\phi+\phi^2$. High peaks are amplified much more than low ones and deep through can
become local maxima.}}
\end{center}
\end{figure}

To extract this information, the Press-Schechter (PS) formalism [25] needs to be extended
to non-gaussian initial conditions.
The PS is an analytical model to calculate the {\it mass function} (that is the
number of object per unit mass at a given redshift per unit volume), within
an appropriate theoretical model (see R. Sheth and S. Shandarin  contributions
in this volume).
The key ingredient is the probability density function (PDF) for the smoothed
dark matter field ${\cal P}(\delta_M)$; in fact, the number density of objects above a given mass $M$ (corresponding to a smoothing
radius $R$) at a given redshift (the {\it mass function})is proportional to the
quantity $P_{>\delta_c}$:
\be
P_{>\delta_c}(\delta_M)=\int_{\delta_c}^{\infty}{\cal P}(\delta_M)d \delta_M
\ee
where $\delta_c$ is the threshold overdensity for the object to collapse,
$P_{>\delta_c} (\delta_M)$ is evidently a function of the redshift of
formation (or collapse) of the object $z_c$, this redshift dependence is
enclosed in $\delta_c$: $\delta_c(z)=\Delta_c/D(z)$. Here $D(z)$ is the
linear growth factor, which in turn depends on the background cosmology, and
$\Delta_c$ is the linear extrapolation of the overdensity for spherical
collapse; $\Delta_c$ is traditionally taken to be $\sim 1.68$ but other values
have also been used (see S. Shandarin and R. Sheth contributions).

For gaussian fields, ${\cal P}(\delta_M)$  is of course well known, but
needs to be computed for non-gaussian initial condition.

Given a physically motivated parameterization of primordial non gaussianity,  
we set off to calculate the PDF for the {\it smoothed} dark matter field
analytically. Of course one could evaluate the PDF from numerical simulations, but this
approach is plagued by the difficulty of properly accounting for the
non-linear way in which resolution and finite box-size effects propagate into
the statistical properties of the non-gaussian field \footnote{For example
imagine computing the power spectrum of a non-gaussian field
$\psi=\phi+\phi^2-\langle\phi^2\rangle$ where $\phi$ is gaussian with a power-law power spectrum $P_{\phi}$. The power spectrum of $\psi$ involves computing the convolution of two $P_{\phi}$. This convolution is an integral
over $k$ from 0 to infinity. When performing the operation on a simulation box
the final result would be as if the integral was truncated at $k\gap 2\pi/L$
and $k\lap 2\pi/l$ where $L$ is the side of the box and $l$ is the grid
resolution.}.

As before we parameterize the non-gaussianity as in equation \ref{ngmapping} where
$\Phi$ can be the potential or the density field.
To properly deal with the smoothing we use a path-integral approach in the
calculation of the PDF:
\be
{\cal P}(\delta_R)=\left\langle \delta^D\left[\phi_R(x)+\epsilon\int d^3y F_R(|x-y|)\phi^2(y)-C-\delta_R(x) \right]\right\rangle
\ee
where the $R$ subscript denotes the smoothed quantity, $\delta^D[\ldots]$ the Dirac
delta function, $\langle \ldots \rangle$ the ensemble average and $F$ is defined
through its Fourier transform ($\tilde{F}_R(k)=W_R(k)T(k)g(k)$ with $W_R$ the
smoothing, $T$ the transfer function and $g=1$ for the density or
$g=-2/3(k/H_0)^2\Omega_{0,m}^{-1}$ for the potential) . In $\phi_R$
smoothing and transfer function are easily accounted for, but in the
non-gaussian part, the presence of the smoothing in $F_R(|x-y|)$, makes the
quantity non-local.
The Dirac delta function can be expressed in its integral representation and
the ensemble average can be written as an integral over all
$\phi$ configurations weighted by the gaussian probability density functional. In this way we
can express an unknown quantity in terms of all known quantities and we are
able to obtain the non-gaussian PDF for the smoothed field analytically. The details of the derivation can be
found in [26].
The main result is that, for mildly non-gaussian initial conditions with small
positive skewness $S_{3,R}$, the
threshold for collapse $\delta_c$ is lowered, in particular [26]:
\be
\delta_c(z_c)\longrightarrow\delta_c(z_c)\left[1-\frac{S_{3,R}}{3}\delta_c(z_c)\right]
\ee
where $S_{3,R}$ is directly proportional to the non-gaussianity parameter
$\epsilon$ of equation \ref{ngmapping}. By lowering the threshold for collapse
rare objects will form more easily, and this has a huge impact on the tails of
the distribution as shown in figures 5 and 6.

\begin{figure}[h]
\begin{center}
\begin{picture}(160,160)
\includegraphics{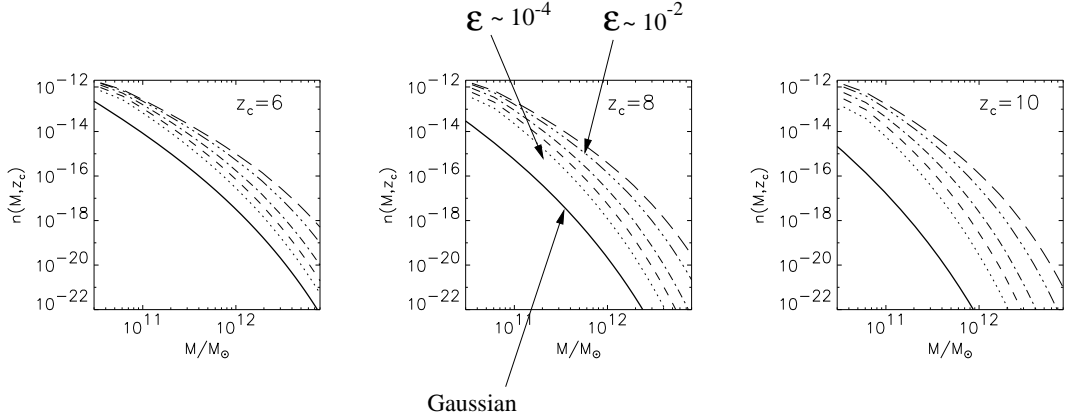}
\end{picture}
\label{fig:higredmodela}
\caption[]{{\it Effects of the non-gaussianity of equation \ref{ngmapping} applied on the
density fluctuation field. The figure shows only galaxy mass scales: the solid line is the mass function for gaussian
initial conditions, all the other lines have non-gaussianity parameter
$10^{-4}\lap\epsilon\lap 10^{-2}$.  For $z_c=8$, a small non-gaussianity
parameter $\epsilon \sim 10^{-4}$ increases the mass function by two orders of
magnitude at about $M \sim 10^{10} M_{\odot}$.}}
\end{center}
\end{figure}

\begin{figure}[h]
\begin{center}
\begin{picture}(160,160)
\includegraphics{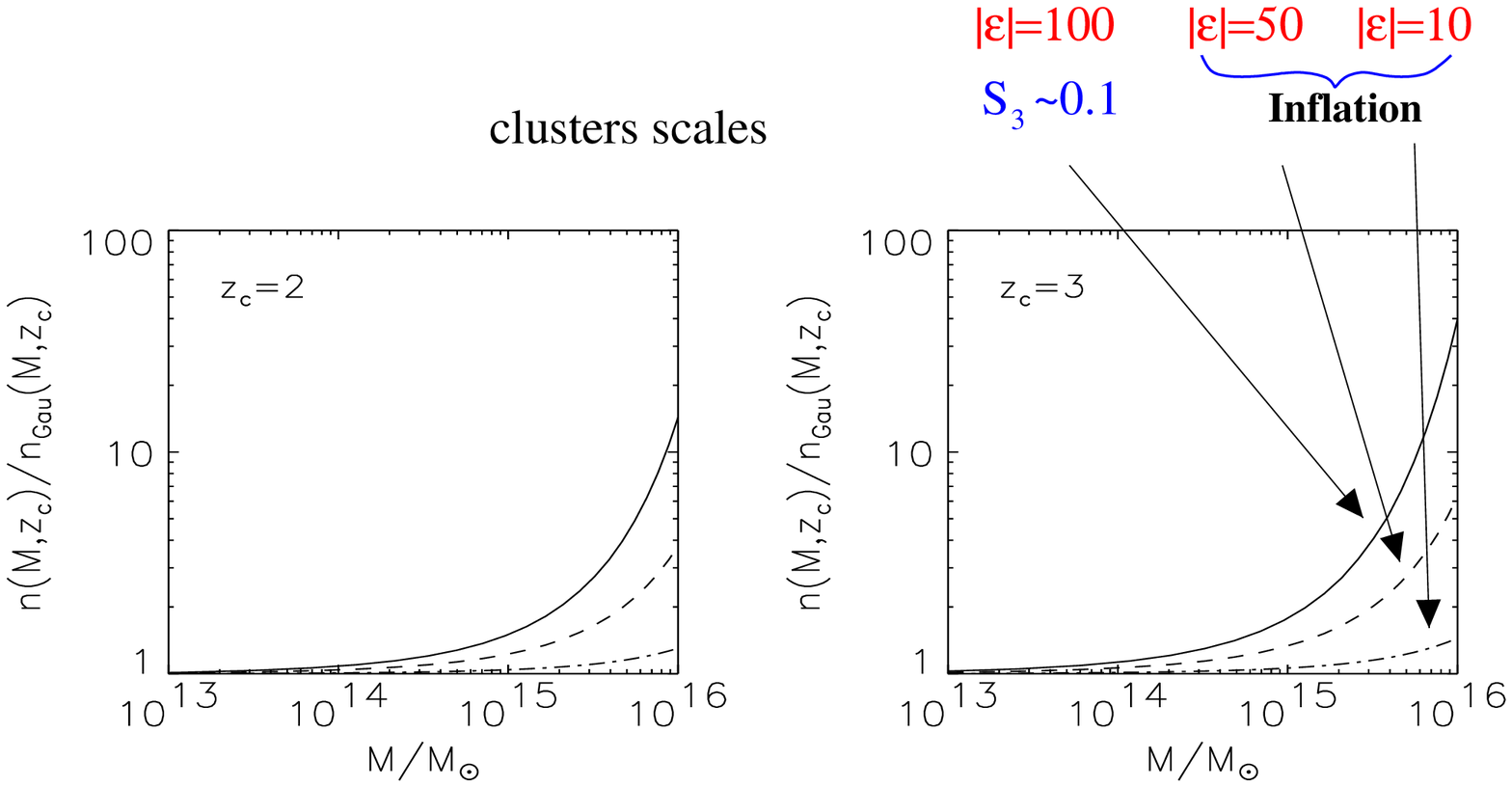}
\end{picture}
\label{fig:higredmodelb}
\caption[]{{\it The non gaussianity of equation \ref{ngmapping} applied to the
gravitational potential field has dramatic effects on cluster scales. In
particular observations of clusters with  $z_c\sim 2-3$ and $M > 10^{15}M_{\odot}$ could
put  constraints on inflationary models. The absolute value for
$\epsilon$ here is relatively big, but the deviation from gaussianity is still
small: the skewness here is of the same order of magnitude as in figure 5.}}
\end{center}
\end{figure}

In particular figure \ref{fig:higredmodela} shows that the non-gaussianity of
equation \ref{ngmapping} applied to the density field  with $\epsilon \sim 10^{-4}$, can
change the number density of objects of mass $M\sim 10^{11}M_{\odot}$ that
collapse at redshift $z_c=8$, by two orders of magnitude.
Conversely, it is clear from figure \ref{fig:higredmodelb} that non-gaussian mapping \ref{ngmapping}, applied to the potential
field, has dramatic effects on cluster scales.
Observations of clusters with  $z_c\gap 2$ and $M\gap 10^{15}M_{\odot}$ could
put some constraints on inflationary models [27].

\subsubsection{A worked example} 
Up to date 6 galaxies with confirmed spectroscopic redshifts have been
observed with redshifts  $5<z<7$. The observed comoving density $N$ is
 for a $\Omega_{0,m}=0.3\;,\;\;\Lambda=0.7$ ($\Lambda$CDM) Universe is $N\geq 8.3\times
10^{-4}(h^{-1}\mbox{Mpc})^{-3}$,[$N\geq 3.6 \times
10^{-4}(h^{-1}\mbox{Mpc})^{-3}$ for an Einstein-de Sitter Unverse].
Their masses are very uncertain, but some estimate can be obtained with simple
arguments about their observed star formation rate, these estimates can then
be compared with Ly$_{\alpha}$ width observations [26].
The gaussian $\Lambda$CDM model predicts $N\geq 5.2\times 10^{-5}(h^{-1}\mbox{Mpc})^{-3}$, a factor $\sim 20$ fewer objects,
while the Einstein-de Sitter model predicts $N\geq
10^{-7}(h^{-1}\mbox{Mpc})^{-3}$: a factor $10^4$ fewer objects! 
Only $\epsilon \sim 10^{-3}$ in the density is needed to reconcile $\Lambda$CDM
predictions with observations. Alternatively, this discrepancy could be
explained postulating an error of about a factor 4 in the mass determination.
As larger telescopes such as NGST get on line it will be possible to determine
masses more accurately and thus constrain the amount of primordial
non-gaussianity on galaxy scales.

\section{Conclusions}
We have shown that non-gaussianity does not necessarily mean non-linearity,
but it is possible to distinguish different kinds of non-linearity: e.g. bias,
gravitational evolution, redshift space distortions.
For the physically motivated  non-gaussian models we considered, it turns out
that CMB bispectrum is better that LSS bispectrum to detect primordial
non-gaussianity: if the future CMB missions will produce maps that
are consistent with the gaussian hypothesis, any non-gaussianity seen in the
LSS bispectrum can be unambiguously attributed to the effects of
non-linearities. Thus, if this is the case, from on-going LSS surveys such as
SDSS and 2dF we will be able to know the bias with few \% accuracy.
We have also shown that,
to measure the bias parameter, ongoing 3D surveys are much better that 2D
ones, even with full sky coverage, but the
method developed has applications in different areas such as CMB and gravitational lensing studies.
To conclude, we have seen different ways to disentangle primordial
non-gaussianity from effects of non-linearity: CMB bispectrum, LSS trispectrum
and the abundance of high-redshift objects such as galaxies and clusters.
These methods probe the Universe a different scales and at different times and
in addition to that they are sensitive to different moments of the
distribution. 
We should therefore conclude that these methods are complementary rather than mutually exclusive.

\vspace*{1cm}

\begin{small}
{\bf Acknowledgments}

I would like to thank my collaborators in this work Alan Heavens, Sabino
Matarrese, Marc Kamionkowski, Limin Wang and Raul Jimenez. 

I also would like
to thank the organizers for a very enjoyable workshop. 
\end{small}

\end{document}